\documentclass[11pt]{article}
\usepackage{longtable}
\usepackage{ae,aecompl,pslatex}
\usepackage[dvips]{graphicx}
\usepackage{enumerate,color}
\usepackage{fullpage}
\usepackage{lineno}

\usepackage{placeins}
\usepackage{psfrag,epsfig,graphicx,mathtools, color,array,dcolumn,tabularx,delarray,longtable,indentfirst,amsmath,amsfonts,amssymb,amsthm,eucal,bbm}
\theoremstyle{bkaexa} 
\usepackage{caption}
\usepackage{setspace}
\theoremstyle{bkaexa} 

\theoremstyle{bkathm} 

\theoremstyle{bkathm} 
\newtheorem{Thm}{Theorem}
\theoremstyle{bkathm} 
\newtheorem{Cor}{Corollary}
\theoremstyle{bkathm} 

\theoremstyle{definition}

\begin{document}
\setstretch{1.5}
\title{Asymptotic expected sensitivity function and its applications to nonparametric correlation estimators}
\author{\normalsize Qingyang Zhang\\
\normalsize Department of Mathematical Sciences, University of Arkansas, AR 72701\\
\normalsize Email: qz008@uark.edu
}
\date{}
\maketitle

\begin{abstract}
We introduce a new type of influence function, the asymptotic expected sensitivity function, which is often equivalent to but mathematically more tractable than the traditional one based on the G\^{a}teaux derivative. To illustrate, we study the robustness of some important rank correlations, including Spearman's and Kendall's correlations, and the recently developed Chatterjee's correlation.
\end{abstract}


\section{Introduction}
In the literature of robust statistics, various measures of robustness have been proposed, including influence function, sensitivity function and breakdown point \cite{Huber, Hampel74, Hampel86}. These measures provide a means to assess the robustness of statistical functionals and estimators, particularly in the face of outliers or deviations from the underlying model. Recently, these robustness measures have found applications in machine learning as well \cite{Koh, ShuZhu, ZhangZhang}. For instance, by quantifying the impact of individual training data points on a model's predictions, these robustness measures may provide insights into how specific samples influence the model's decision-making process. This capability enables the identification of critical data points that significantly affect the model's performance, facilitating more informed data curation and model debugging.

In this work, we focus on two key measures: the influence function for functionals and sensitivity function for estimators. We begin with the definitions and a brief discussion on their respective limitations. For random variable $X$ with cumulative distribution function (\textit{c.d.f.}) $F$, the influence function ($IF$) of a statistical functional $R(F)$ at $X = x$ is defined as 
\begin{equation*}
\mbox{IF}(x,~ R,~ F) = \lim_{\epsilon\downarrow 0}\frac{R[(1-\epsilon)F + \epsilon\Delta_{x}] - R(F)}{\epsilon},
\end{equation*}
where $\Delta_{x}$ is a Dirac distribution which puts all its mass on $X = x$. The influence function $\mbox{IF}(x,~ R,~ F)$ is essentially a G\^{a}teaux derivative at $X=x$, which quantifies the sensitivity of $R(F)$ to infinitesimal modifications in a single data point $x$. A substantial influence function indicates that the functional is highly susceptible, whereas a modest influence function suggests that the functional is more robust to $X = x$. For all linear functionals, i.e., functionals that can be written as $R(F) = \int g(t)dF(t)$, the influence function is $g(x) - R(F)$.

Sensitivity function (\textit{SF}), sometimes also called empirical influence function, can be viewed as a sample-based counterpart of influence function. Let $F_{n}$ be the empirical distribution function, and $R_{n}$ be an estimator of $R$ based on \textit{i.i.d.} samples $X_{1},~...,~X_{n}$. An estimator $R_{n}$ is called a plug-in estimator if $R_{n} = R(F_{n})$. The SF of $R_{n}$ at $X=x$ is defined as 
\begin{equation*}
\mbox{SF}(x,~ R_{n},~ F_{n}) = (n+1)\left[ R_{n+1}(X_{1},~...,~X_{n},~x) - R_{n}(X_{1},~...,~X_{n}) \right ],
\end{equation*}
which represents the change in estimator when a new observation $x$ is added to the existing sample of size $n$. The asymptotic version of sensitivity function (\textit{ASF}) is defined as  $\lim_{n\rightarrow\infty}\mbox{SF}(x,~ R_{n},~ F_{n})$. It is noteworthy that the above definitions of \textit{IF}, \textit{SF} and \textit{ASF} are extendable to any dimension. 

Though the influence function and sensitivity function have been widely used as measures of robustness, they both have certain limitations from practical point of view. For instance, the influence function is based on G\^{a}teaux derivative, which can be difficult to derive for some complicated functionals. Consider Dette-Siburg-Stoimenov's dependence measure \cite{dss}, also known as the limit of Chatterjee's rank correlation \cite{chatterjee}, expressed as
\begin{equation*}
\xi(X, Y) = \frac{\int V(\mathbb{E}(\mathbbm{1}\{Y\geq t|X\}))dF_{Y}(t)}{\int V(\mathbbm{1}\{Y\geq t\})dF_{Y}(t)}.
\end{equation*}
This functional, characterized by a ratio of integrals, presents a relatively intricate structure. The robustness of $\xi(X, Y)$ remains unexplored due to the difficulty of deriving its G\^{a}teaux derivative. 

The sensitivity function is generally more straightforward to derive, but it relies on the random samples, making it not suitable as a population-level measure. For certain consistent estimators, the asymptotic sensitivity function (obtained by letting sample size approach infinity) are deterministic and equivalent to the influence function. However, this equivalence does not hold for many other functionals and estimators. Notably, Croux (1998) demonstrated that the asymptotic sensitivity function of the sample median, rather than being deterministic, follows an exponential-type distribution \cite{croux}. 

Motivated by both Croux's work and Dette-Siburg-Stoimenov's dependence measure, we propose a new robustness measure based on the asymptotic expectation of sensitivity function, termed asymptotic expected sensitivity function (\textit{AESF}). For many functionals, the proposed measure can be derived similarly to sensitivity function, without requiring the G\^{a}teaux derivative. Crucially, unlike the asymptotic sensitivity function, the \textit{AESF} is always deterministic, and we demonstrate that in many cases, it coincides with the influence function.

The remainder of the paper is structured as follows: Section 2 introduces the new robustness measure with some elementary examples, and then discusses its connections to \textit{IF}, \textit{SF}. In Section 3, we apply the proposed measure to analyze the robustness of three important rank correlations, including Spearman's, Kendall's and Chatterjee's correlations. Section 4 discusses and concludes the paper. 

\section{A new influence function}
In this section, we define the new influence function \textit{AESF}, followed by some illustrative examples. We then briefly discuss the relationships between \textit{AESF}, \textit{ASF} and \textit{IF}, shedding light on their interconnectedness. For clarity, we focus on univariate case, but it is important to note that the new measure can be readily extended to multivariate settings. 

\subsection{Definition and elementary examples}
Consider the expected sensitivity function, expressed as 
\begin{equation*}
\mbox{ESF}(x,~ R,~ F,~n) = (n+1)\mathbb{E}\left[ R_{n+1}(X_{1},~ ...,~ X_{n}, ~x) - R_{n}(X_{1},~ ..., ~X_{n}) \right].
\end{equation*}
The expectation in \textit{ESF} is to eliminate the randomness of $SF$. In spirit, it is similar to the $SF$ based on stylized sample proposed by Andrews et al. (1972) (page 96, \cite{stylized}), which is essentially a pseudo-sample consisting of the $n$ expected normal order statistics from $(X_{1},~ ...,~ X_{n})$. However, the $ESF$ distinguishes itself with more tractable theoretical properties. 

The asymptotic expected sensitivity function (\textit{AESF}) can be then defined as 
\begin{equation*}
\mbox{AESF}(x,~ R,~ F) = \lim_{n\rightarrow\infty}\mbox{ESF}(x,~ R,~ F,~n).
\end{equation*}

It is not hard to see that \textit{AESF} is always deterministic. While $AESF$ often coincides with the expectation of \textit{ASF} for numerous functionals and their estimators, their general equivalence is not guaranteed, unless for uniformly bounded sensitivity function where the Lebesgue's Dominated Convergence Theorem applies. We use the following elementary examples to elucidate the derivation of \textit{AESF} and its relationship to \textit{ASF} and \textit{IF}.

\subsubsection*{Example 1: Mean}
The expectation functional, $\mu(F) = \int_{-\infty}^{\infty} t dF(t)$, is a linear functional and its influence function is $\mbox{IF}(x,~ \mu,~ F) = x-\mu$, which is unbounded. The sample mean can be treated as a plug-in estimator of $\mu(F)$, i.e., $\mu_{n} = \mu(F_{n}) = \sum_{i=1}^{n}X_{i}/n$. The sensitivity function of $\mu_{n}$ is 
\begin{align*}
\mbox{SF}(x,~ \mu_{n},~ F_{n}) & = (n+1)\left[ \frac{x+\sum_{i=1}^{n}X_{i}}{n+1} - \frac{\sum_{i=1}^{n}X_{i}}{n} \right] \\
& = x - \frac{\sum_{i=1}^{n}X_{i}}{n}. 
\end{align*}

By Strong Law of Large Numbers (SLLN), $\sum_{i=1}^{n}X_{i}/n$ converges to $\mu$ almost surely, therefore for any given $x$, $\mbox{SF}(x,~ \mu_{n},~ F_{n})$ converges to $x - \mu$ almost surely.

The expected sensitivity function is
\begin{align*}
\mbox{ESF}(x,~ \mu,~ F,~n)  & = \mathbb{E}[\mbox{SF}(x, ~R_{n}, ~F_{n})]\\
 & = x - \frac{\sum_{i=1}^{n}E(X_{i})}{n} \\
& = x - \mu,
\end{align*}
which does not depend on $n$, and $\mbox{AESF}(x,~ \mu,~ F) = x - \mu$. Therefore, we have $\mbox{AESF}(x,~ \mu,~ F) = \mbox{IF}(x,~ \mu,~ F)$, and $\mbox{SF}(x, ~\mu_{n}, ~F_{n})$ converges almost surely to $\mbox{AESF}(x,~ \mu,~ F)$ for any $x$.

\subsubsection*{Example 2: Variance}
The variance functional, $\sigma^2(F) = \int_{-\infty}^{\infty} [t-\int_{-\infty}^{\infty} s dF(s)]^2 dF(t)$, has the influence function $\mbox{IF}(x,~ \sigma^2,~ F) = (x-\mu)^2 - \sigma^2$, which is unbounded.

The sensitivity function of sample variance $\sigma^2_{n}$ is
\begin{align*}
\mbox{SF}(x,~ \sigma^2_{n},~ F_{n}) & = (n+1)\left[ \frac{x^2 + \sum_{i=1}^{n}X^2_{i}}{n+1} - \left(\frac{x+\sum_{i=1}^{n}X_{i}}{n+1}\right)^2 -  \frac{\sum_{i=1}^{n}X^2_{i}}{n} + \left(\frac{\sum_{i=1}^{n}X_{i}}{n}\right)^2 \right] \\
& = \frac{n}{n+1}x^2 - \frac{\sum_{i=1}^{n}X_{i}}{n}\frac{2n}{n+1}x + \frac{2n+1}{n+1}\left( \frac{\sum_{i=1}^{n}X_{i}}{n} \right)^2 - \frac{\sum_{i=1}^{n}X^2_{i}}{n}.
\end{align*}
By SLLN, for any $x$, $\mbox{SF}(x, \sigma^2_{n}, F_{n})$ converges to $(x-\mu)^2 - \sigma^2$ almost surely.

The expected sensitivity function is 
\begin{align*}
\mbox{ESF}(x,~ \sigma^2,~ F,~n)  & = \mathbb{E}\left[\mbox{SF}(x,~ \sigma^2_{n},~ F_{n})\right]\\
 & =  \frac{n}{n+1}x^2 - \frac{2n}{n+1}x\mu + \frac{n}{n+1}\mu^2 -\frac{n^2-n-1}{n(n+1)}\sigma^2, 
\end{align*}
and $\mbox{AESF}(x,~ \sigma^2,~ F) = (x-\mu)^2 - \sigma^2$. Therefore we have $\mbox{AESF}(x,~ \sigma^2,~ F) = \mbox{IF}(x,~ \sigma^2,~ F)$, and $\mbox{SF}(x, ~\sigma^2_{n}, ~F_{n})$ converges almost surely to $\mbox{AESF}(x,~ \sigma^2,~ F)$ for any $x$.

\subsubsection*{Example 3: Maximum of uniform distribution}
Let $F$ be a uniform distribution on $[0, \theta]$, we consider a simple functional $R(F) = \inf\{t|F(t) = 1\} = \theta$, and the plug-in estimator $R_{n} = R(F_{n}) = \inf\{t|F_{n}(t) = 1\} = X_{(n)}$, where $X_{(1)}\leq ... 
\leq X_{(n)}$. For any $0\leq\epsilon<1$ and $0\leq x\leq \theta$, we have $R[(1-\epsilon)F + \epsilon\Delta_{x}] = \theta$, thus 
\begin{equation*}
\mbox{IF}(x,~ R,~ F) = 0 \text{~~for~~} 0\leq x\leq \theta.
\end{equation*}

The sensitivity function of $R_{n}$ is 
\begin{equation*}
\mbox{SF}(x,~ R_{n},~ F_{n}) = (n+1)(x-X_{(n)})\mathbf{1}\left[X_{(n)}<x\right].
\end{equation*}

As $X_{(n)}\xrightarrow[]{a.s.}\theta$, for any $x<\theta$, $\mbox{SF}(x,~ R_{n},~ F_{n})\xrightarrow[]{a.s.}0$. For $x = \theta$, the \textit{c.d.f.} of $\mbox{SF}(x,~ R_{n},~ F_{n})$ is
\begin{align*}
\mbox{P}\left[\mbox{SF}(x,~ R_{n},~ F_{n})<s \right] & = 1 - \mbox{P}\left[ X_{(n)} < \theta - \frac{s}{n+1} \right] \\
& = 1 - \left[ 1-\frac{s}{\theta(n+1)} \right]^{n},
\end{align*}
which converges in distribution to an exponential distribution with rate parameter $\theta$, i.e., the \textit{ASF} is not deterministic.

Furthermore, because the \textit{c.d.f.} of $X_{(n)}$ is $F_{X_{(n)}}(x) = (x/\theta)^{n}$ for $0\leq x\leq \theta$, we have
\begin{align*}
\mbox{ESF}(x,~ R,~ F,~ n) & = \mathbb{E}\left[ (n+1)(x-X_{(n)})\mathbf{1}\left(X_{(n)}<x\right) \right] \\
& =  \mathbb{E}\left[ x(n+1)\mathbf{1}\left(X_{(n)}<x\right) \right]  -  \mathbb{E}\left[ (n+1)X_{(n)}\mathbf{1}\left(X_{(n)}<x\right) \right] \\ 
& = x(n+1)\left(\frac{x}{\theta}\right)^{n} - xn\left(\frac{x}{\theta}\right)^{n} \\
& = x\left(\frac{x}{\theta}\right)^{n},
\end{align*}
and
\begin{equation*}
\mbox{AESF}(x,~ R,~ F) = \begin{cases}
0, & 0\leq x<\theta \\
\theta, & x = \theta.
\end{cases}
\end{equation*}
From the derivations above, it can be seen that the $IF$, $AESF$ and $ASF$ coincide for $0\leq x < \theta$, but they diverge notably at $x=\theta$. Specifically, we observe the following
\begin{itemize}
\item $\mbox{IF}(x = \theta,~ R,~ F) = 0$
\item $\mbox{AESF}(x = \theta,~ R,~ F) = \theta$
\item $\lim_{n\rightarrow\infty}\mbox{SF}(x = \theta,~ R_{n},~ F_{n})$ is not deterministic.
\end{itemize}
This example highlights the potential for significant discrepancies between these metrics for certain nonlinear functionals and specific values of $x$. The subsequent section will explore the relationships between these interconnected metrics and establish sufficient conditions under which they achieve equivalence.

\subsection{Relationships between \textit{AESF}, \textit{SF} and \textit{IF}}
For any linear functional $R(F) = \int g(t) dF(t)$ with $|\mathbb{E}[g(X)]|<\infty$, let $R_{n}$ be the plug-in estimator, i.e., $R_{n} = \int g(t) dF_{n}(t) = \sum_{i = 1}^{n}g(X_{i})/n$, where $X_{1},~ ...,~ X_{n}$ are $i.i.d.$ samples, we have
\begin{align*}
\mbox{IF}(x,~ R,~ F) & = g(x) -  \int g(t) dF(t),\\
\mbox{SF}(x, ~R_{n}, ~F_{n}) & = g(x) - \frac{1}{n}\sum_{i = 1}^{n}g(X_{i}).
\end{align*}
As $\sum_{i = 1}^{n}g(X_{i})/n \xrightarrow[]{a.s.} \int g(t) dF(t)$, we have $\mbox{SF}(x, ~R_{n}, ~F_{n}) \xrightarrow[]{a.s.} g(x) - \int g(t) dF(t) = \mbox{IF}(x,~ R,~ F)$. In addition, $\mbox{ESF}(x,~ R,~ F,~n) = g(x) - \sum_{i = 1}^{n}\mathbb{E}\left[g(X_{i})\right]/n  = g(x) -  \int g(t) dF(t) = \mbox{IF}(x,~ R,~ F)$, we have for any $x$
$$\mbox{SF}(x, ~R_{n}, ~F_{n}) \xrightarrow[]{a.s.} \mbox{AESF}(x,~ R,~ F) = \mbox{IF}(x,~ R,~ F).$$

Because the expectation functional $\mu(F)$ is linear, we have $\mbox{SF}(x, ~\mu_{n}, ~F_{n}) \xrightarrow[]{a.s.} \mbox{AESF}(x,~ \mu,~ F) = \mbox{IF}(x,~ \mu,~ F)$ for any $x$. The following theorem extends the results for linear functionals to any continuous and bounded transformation (a detailed proof is provided in the Appendix). 

\begin{Thm}
Consider a statistical functional $R(F) = \phi\left[\int g(t) dF(t)\right]$ and its plug-in estimator $R_{n} = \phi\left[\int g(t) dF_{n}(t)\right]$. Suppose $\int |g(t)| dF(t)<\infty$, $\phi$ is differentiable with continuous and bounded $\phi^{'}$, then for any $x$, we have
$$\mbox{SF}(x, ~R_{n}, ~F_{n}) \xrightarrow[]{a.s.} \mbox{AESF}(x,~ R,~ F) = \mbox{IF}(x,~ R,~ F) = \left[g(x) - \int g(t) dF(t)\right]\phi^{'}\left[\int g(t) dF(t)\right].$$
\end{Thm}

The following corollary can be easily obtained by a linear combination:
\begin{Cor} 
If $[\phi_{l}(\cdot),~ g_{l}(\cdot)]_{l = 1,~ ...,~ L}$ all satisfy the conditions in Theorem 1, $(a_{1},~ ...,~ a_{L})$ are constants, then for functional $R(F) = \sum_{l = 1}^{L}a_{l}\phi_{l}\left[\int g_{l}(t) dF(t)\right]$, we have 
$$\mbox{SF}(x, ~R_{n}, ~F_{n}) \xrightarrow[]{a.s.} \mbox{AESF}(x,~ R,~ F) = \mbox{IF}(x,~ R,~ F) = \sum_{l=1}^{L}a_{l}\left[g_{l}(x) - \int g_{l}(t) dF(t)\right]\phi_{l}^{'}\left[\int g_{l}(t) dF(t)\right].$$
\end{Cor}
By Corollary 1, it is immediate that for variance, we have $\mbox{SF}(x, ~\sigma^2_{n}, ~F_{n}) \xrightarrow[]{a.s.} \mbox{AESF}(x,~ \sigma^2,~ F) = \mbox{IF}(x,~ \sigma^2,~ F)$ for any $x$, simply by letting $a_{1} = 1,~ a_{2} = -1,~ g_{1}(t) = t^2,~ g_{2}(t) = t,~ \phi_{1}(z) = z,~ \phi_{2}(z) = z^2$. 

For functionals that cannot be expressed as continuous transformations of linear functionals, equivalence results are not generally guaranteed. An open question for this is the identification of sufficient and necessary conditions for such equivalence.

\section{Application to some rank correlations}
In this section, we derive the $AESF$ of some important nonparametric correlations, including the widely used Kendall's and Spearman's correlations and a recent one developed by Chatterjee (2021). It is found that for Kendall's and Spearman's correlations, the $AESF$ is same as the $IF$. For Chatterjee's correlation, the $IF$ is unknown due to the complicated form of its functional representation, a.k.a., Dette-Siburg-Stoimenov's dependence measure, but the derivation of $AESF$ is tractable, and we use $AESF$ as an alternative metric to study its robustness under various correlative patterns.

\subsection{Kendall's and Spearman's correlations}
Let $(X_{1},~ Y_{1}),~ ...,~ (X_{n},~ Y_{n})$ be $n$ i.i.d. samples of $(X,~ Y)$. The Kendall's correlation is defined as 
\begin{equation*}
\tau_{n} =  \frac{2}{n(n-1)}\sum_{i<j} \mbox{sgn}\left[ (X_{i} - X_{j})(Y_{i} - Y_{j}) \right],
\end{equation*}
where $\mbox{sgn}(t)$ equals 1 for $t\geq 0$ and -1 for $t<0$. The functional representation of $\tau_{n}$ is 
\begin{equation*}
\tau =  \mathbb{E}\left\{\mbox{sgn} [(X - X^{'})(Y - Y^{'})] \right\},
\end{equation*}
where $(X^{'}, Y^{'})$ represents an independent copy of $(X, Y)$. It can be seen that Kendall's correlation is a U-statistic, and its sensitivity function is 
\begin{align*}
\mbox{SF}[(x,~ y), ~\tau_{n}, ~F_{n}] & = (n+1)\left\{ \frac{2}{n(n+1)}\sum_{i<j} \mbox{sgn}\left[ (X_{i} - X_{j})(Y_{i} - Y_{j}) \right] + \frac{2}{n(n+1)}\sum_{i=1}^{n} \mbox{sgn}\left[ (X_{i} - x)(Y_{i} - y) \right] - \tau_{n} \right\} \\
& = \frac{2}{n}\sum_{i=1}^{n} \mbox{sgn}\left[ (X_{i} - x)(Y_{i} - y) \right] -  \frac{4}{n(n-1)}\sum_{i<j} \mbox{sgn}\left[ (X_{i} - X_{j})(Y_{i} - Y_{j}) \right]. 
\end{align*}
Note that $\mbox{sgn}(t) = 2\cdot\mathbf{1}(t\geq 0) - 1$, we have 
\begin{align*}
\mathbb{E}\left\{ \mbox{SF}[(x,~ y), ~\tau_{n}, ~F_{n}] \right\} & =  4P[(X - x)(Y - y)>0] - 2 - \mathbb{E}\left\{\mbox{sgn} [(X - X^{'})(Y - Y^{'})] \right\} \\
& = 4P[(X - x)(Y - y)>0] - 2 - \tau.
\end{align*}
The $ESF$ does not rely on sample size $n$, therefore 
\begin{equation*}
\mbox{AESF}[(x,~ y),~ \tau,~ F] = 4P[(X - x)(Y - y)>0] - 2 - \tau,
\end{equation*}
which is equivalent to the $\mbox{IF}[(x,~ y),~ \tau,~ F]$ (Proposition 1, \cite{crouxdehon}), and uniformly bounded by 3. As a special case, for bivariate normal distribution with correlation $\rho$, we have 
\begin{equation*}
\mbox{AESF}[(x,~ y),~ \tau,~ F] = 8\Phi_{\rho}(x, y) - 4\Phi(x) - 4\Phi(y) + 2 - \frac{2}{\pi}\arcsin(\rho).
\end{equation*}
Figure 1 below demonstrates the $AESF$ values for a standard bivariate normal distribution with $\rho = 0.7$.

\begin{center}
[Figure 1 about here]
\end{center}

Next we derive the $AESF$ for Spearman's correlation. The correlation estimator is 
\begin{equation*}
S_{n} = 1-\frac{6\sum_{i=1}^{n}[r(X_{i}) - r(Y_{i})]^2}{n(n-1)(n+1)},
\end{equation*}
where $r(X_{i}) = \sum_{j=1}^{n}\mathbb{I}(X_{j}\leq X_{i})$ and $r(Y_{i}) = \sum_{j=1}^{n}\mathbb{I}(Y_{j}\leq Y_{i})$. The functional representation of $S_{n}$ is 
\begin{equation*}
S = 12\mathbb{E}[F_{X}(X)F_{Y}(Y)] - 3.
\end{equation*}
Let $r^{*}(X_{i})$ and $r^{*}(Y_{i})$ be the ranks of $X_{i}$ and $Y_{i}$ after $(x, y)$ is added, the $SF$ can be decomposed as follows
\begin{equation*}
\mbox{SF}[(x,~ y), ~S_{n}, ~F_{n}] = \Delta_{n1} + \Delta_{n2} + \Delta_{n3}, 
\end{equation*}
where 
\begin{align*}
\Delta_{n1}  & = \frac{ 6\sum_{i=1}^{n}[r(X_{i}) - r(Y_{i})]^2 - 6\sum_{i=1}^{n}[r^{*}(X_{i}) - r^{*}(Y_{i})]^2 }{n(n+2)}, \\
\Delta_{n2}  & = -\frac{6[r^{*}(x) - r^{*}(y)]^2}{n(n+2)}, \\
\Delta_{n3}  & = \frac{18\sum_{i=1}^{n}[r(X_{i}) - r(Y_{i})]^2}{n(n-1)(n+2)}.
\end{align*}
First, for $\Delta_{n2}$, we have 
\begin{equation*}
\Delta_{n2} = \frac{12r^{*}(x)r^{*}(y)}{(n+1)^2} + O(1/n).
\end{equation*}
By Glivenko-Cantelli Theorem
\begin{equation*}
\lim_{n\rightarrow \infty}\Delta_{n2} = 12F_{X}(x)F_{Y}(y), ~a.s.
\end{equation*}
In addition, as $[r^{*}(x) - r^{*}(y)]^2\leq (n-1)^2$, $\Delta_{n2}$ is uniformly bounded by 6. By Lebesgue's Dominated Convergence Theorem
\begin{equation*}
\lim_{n\rightarrow \infty}\mathbb{E}(\Delta_{n2}) =  \mathbb{E}(\lim_{n\rightarrow \infty}\Delta_{n2}) = 12F_{X}(x)F_{Y}(y).
\end{equation*}
For $\Delta_{n3}$, it is easy to see that 
\begin{equation*}
\Delta_{n3} = \frac{3(n+1)}{n+2}(1-S_{n}), ~\text{and}~ 0<\Delta_{n3}<6.
\end{equation*}
By Lebesgue's Dominated Convergence Theorem
\begin{equation*}
\lim_{n\rightarrow \infty}\mathbb{E}(\Delta_{n3}) =  \mathbb{E}(\lim_{n\rightarrow \infty}\Delta_{n3}) = 3(1-S).
\end{equation*}
Last for $\Delta_{n1}$, we have 
\begin{align*}
\Delta_{n1} & = \frac{12\sum_{i=1}^{n}[r^{*}(X_{i})r^{*}(Y_{i}) - r(X_{i})r(Y_{i})]}{n(n+2)} - 12 + O(1/n) \\
& = \frac{12}{n(n+2)}\sum_{i=1}^{n}r(X_{i})\mathbf{1}(Y_{i}\geq y) + \frac{12}{n(n+2)}\sum_{i=1}^{n}r(Y_{i})\mathbf{1}(X_{i}\geq x) - 12 + O(1/n).
\end{align*}
Therefore
\begin{align*}
\mathbb{E}(\Delta_{n1}) & = 12\mathbb{E}[F_{X}(X)\mathbf{1}(Y\geq y)] + 12\mathbb{E}[F_{Y}(Y)\mathbf{1}(X\geq x)] -12 + O(1/n), \\
\lim_{n\rightarrow\infty}\mathbb{E}(\Delta_{n1}) & = 12\mathbb{E}[F_{X}(X)\mathbf{1}(Y\geq y)] + 12\mathbb{E}[F_{Y}(Y)\mathbf{1}(X\geq x)] -12.
\end{align*}
Summarizing the results above, we have
\begin{equation*}
\mbox{AESF}[(x,~ y),~ S,~ F] = 12F_{X}(x)F_{Y}(y) + 12\mathbb{E}[F_{X}(X)\mathbf{1}(Y\geq y)] + 12\mathbb{E}[F_{Y}(Y)\mathbf{1}(X\geq x)] -3S - 9,
\end{equation*}
which is equivalent to the $\mbox{IF}[(x, y),~ S,~ F]$ (Proposition 1, \cite{crouxdehon}), and uniformly bounded. As a special case, for bivariate normal distribution with correlation $|\rho|<1$, we have 
\begin{align*}
\mbox{AESF}[(x,~ y),~ S,~ F] = & 12\Phi(x)\Phi(y) + 12\left\{\mathbb{E}\left[ \Phi(X)\Phi\left( \frac{\rho X - y}{\sqrt{1-\rho^2}} \right)\right] + \mathbb{E}\left[ \Phi(Y)\Phi\left( \frac{\rho Y - x}{\sqrt{1-\rho^2}} \right)\right] \right\} \\
& - \frac{18}{\pi}\arcsin\left( \frac{\rho}{2} \right)-9.
\end{align*}
Figure 2 below demonstrates the $AESF$ for a standard bivariate normal distribution with $\rho = 0.7$. Figure 3 presents the comparison of Kendall's and Spearman's correlation under the same setting, where it can be seen that Kendall's correlation is more robust (smaller absolute $AESF$) than Spearman's correlation for most of the discordant pairs (red), but less robust to the concordant pairs.  

\begin{center}
[Figure 2 about here]
\end{center}

\begin{center}
[Figure 3 about here]
\end{center}

\subsection{Chatterjee correlation}
Chatterjee (2021) introduced a new correlation test, which is distribution-free, asymptotically normal, and consistent against all alternatives \cite{chatterjee}. Due to its nice properties, Chatterjee's correlation has attracted much attention over the past two years \cite{ShiHan, LinHan, auddy, azadkia.chatterjee, LinHan22, Zhang23a, Zhang23b, CaoBickel, Deb.etal, Huang, ShiHan22, HanHuang, chatterjeenet}. However, the robustness of this new metric remains unexplored, partially because the influence function of its functional representation is difficult to derive. 

Here, we study the robustness of Chatterjee's correlation using the proposed $AESF$. We begin with the definition of this new correlation measure. Let $X$ and $Y$ be two continuous variables, and $(X_{i},~ Y_{i})_{i=1,...,n}$ be n $i.i.d.$ samples. Assuming that $X_{i}$'s and $Y_{i}$'s have no ties, the data can be uniquely arranged as $(X_{(1)}, Y_{(1)}),~ ...,~ (X_{(n)}, Y_{(n)})$, such that $X_{(1)}<\cdots<X_{(n)}$. Here $Y_{(1)},~ ...,~ Y_{(n)}$ denote the concomitants. Let $r_{i}$ be the rank of $Y_{(i)}$, i.e., $r_{i} = \sum_{k=1}^{n}\mathbbm{1}\{Y_{(k)}\leq Y_{(i)}\}$, Chatterjee's correlation $\xi_{n}(X, Y)$ is defined as
\begin{equation}
\xi_{n}(X, Y) = 1-\frac{3\sum_{i=1}^{n-1}|r_{i+1}-r_{i}|}{n^2-1}.
\end{equation}
 
The asymptotic behavior of $\xi_{n}(X, Y)$ and related problems have been examined in recent papers. For instance, in his original paper \cite{chatterjee}, Chatterjee showed that 
\begin{equation*}
\xi_{n}(X, Y)\rightarrow \xi(X, Y) = \frac{\int V(E(\mathbbm{1}\{Y\geq t|X\}))dF_{Y}(t)}{\int V(\mathbbm{1}\{Y\geq t\})dF_{Y}(t)},  ~a.s.
\end{equation*}
The limiting quantity $\xi(X, Y)$ is also known as Dette-Siburg-Stoimenov's dependence measure \cite{dss}, which is between 0 and 1 (0 if and only if $X$ and $Y$ are independent, 1 if and only if $Y$ is a measurable function of $X$). 

Let $(r^{*}_{1}, ..., r^{*}_{n+1})$ be the ranks of concomitants after $(x, y)$ being added, we have
\begin{align*}
\mbox{SF}[(x,~ y),~ \xi_{n},~ F_{n}] & = (n+1)\bigg\{ \xi_{n+1}[(X_{1},~ Y_{1}), ...,~(X_{n},~ Y_{n}),~(x,~ y)] - \xi_{n}[(X_{1},~ Y_{1}),~ ..., ~(X_{n},~ Y_{n})] \bigg\} \\
& = 3(n+1)\left\{ \frac{\sum_{i=1}^{n-1}|r_{i+1}-r_{i}|}{n^2-1} - \frac{\sum_{j=1}^{n}|r^{*}_{j+1}-r^{*}_{j}|}{(n+1)^2-1} \right\} \\
& = \frac{3\sum_{i=1}^{n-1}|r_{i+1}-r_{i}|}{n-1} - \left( 1-\frac{2n+1}{n^2+2n} \right) \frac{\sum_{j=1}^{n}|r^{*}_{j+1}-r^{*}_{j}|}{n-1}.
\end{align*}
Further, let 
\begin{equation*}
\Delta_{n} = \frac{3\sum_{j=1}^{n}|r^{*}_{j+1}-r^{*}_{j}|}{n-1} - \frac{3\sum_{i=1}^{n-1}|r_{i+1}-r_{i}|}{n-1},
\end{equation*}
the sensitivity function can be written as 
\begin{equation}
\mbox{SF}[(x,~ y),~ \xi_{n},~ F_{n}] = \frac{2n+1}{n^2+2n} \frac{3\sum_{i=1}^{n-1}|r_{i+1}-r_{i}|}{n-1} - \Delta_{n} \left( 1- \frac{2n+1}{n^2+2n} \right).
\end{equation}

Let $N(i)$ be the index $j$ such that $X_{j}$ is immediately to the right of $X_{i}$ if we arrange the $X$'s in increasing order. If $X_{i}$ is the rightmost value, define $N(i)$ arbitrarily, which does not matter since the contribution of a single
term in the above sum is $O(1/n)$. For the first term in Equation (2), we have
\begin{equation*}
\lim_{n\rightarrow\infty}\mathbb{E}\left[ \frac{\sum_{i=1}^{n-1}|r_{i+1}-r_{i}|}{n^2} \right] = \lim_{n\rightarrow\infty}\mathbb{E}\left[ \frac{1}{n}\sum_{i=1}^{n}|F_{Yn}(Y_{i}) - F_{Yn}(Y_{N(i)})| \right]. 
\end{equation*}

Let 
\begin{align*}
Q_{n} & = \frac{1}{n}\sum_{i=1}^{n}|F_{Yn}(Y_{i}) - F_{Yn}(Y_{N(i)})|, \\
Q^{'}_{n} & = \frac{1}{n}\sum_{i=1}^{n}|F_{Y}(Y_{i}) - F_{Y}(Y_{N(i)})|, 
\end{align*}
and
\begin{equation*}
\delta_{n} = \sup_{a\in\mathbb{R}} |F_{Yn}(a) - F_{Y}(a)|,
\end{equation*}
we have 
\begin{equation*}
|Q^{'}_{n} - Q_{n}| < 2\delta_{n}.
\end{equation*}
By Glivenko-Cantelli Theorem, $\delta_{n}\rightarrow 0$ almost surely as $n\rightarrow\infty$. Since $0\leq\delta_{n}\leq 1$, we have 
\begin{equation*}
\lim_{n\rightarrow\infty}\mathbb{E}|Q^{'}_{n} - Q_{n}| = 0,
\end{equation*}
and 
\begin{align*}
\lim_{n\rightarrow\infty}\mathbb{E}\left[ \frac{\sum_{i=1}^{n-1}|r_{i+1}-r_{i}|}{n^2} \right] & = \lim_{n\rightarrow\infty}\mathbb{E}\left[ \frac{1}{n}\sum_{i=1}^{n}|F_{Y}(Y_{i}) - F_{Y}(Y_{N(i)})| \right]  \\
& = 1- 2\lim_{n\rightarrow\infty}\mathbb{E}\left[ \frac{1}{n}\sum_{i=1}^{n}\min(F(Y_{i}), F(Y_{N(i)}))\right]. 
\end{align*}

By Lemma (9.10) of \cite{chatterjee}, we have 
\begin{equation*}
\lim_{n\rightarrow\infty}\mathbb{E}\left[ \frac{\sum_{i=1}^{n-1}|r_{i+1}-r_{i}|}{n^2} \right]  = 1 - 2\mathbb{E}_{Y'}\mathbb{E}_{X}\left[ P^{2}(Y\geq Y'|X) \right],
\end{equation*}
and 
\begin{equation*}
\lim_{n\rightarrow\infty}\mathbb{E}\left[ \frac{2n+1}{n^2+2n} \frac{3\sum_{i=1}^{n-1}|r_{i+1}-r_{i}|}{n-1}  \right]  = 6 - 12\mathbb{E}_{Y'}\mathbb{E}_{X}\left[ P^{2}(Y\geq Y'|X) \right],
\end{equation*}
where $Y'$ represents an independent copy of $Y$. Suppose $X_{(m)}< x\leq X_{(m+1)}$, an important observation is 
\begin{equation*}
\Delta_{n} = \frac{3}{n-1}\left[ \sum_{i\neq m}\mathbf{1}\{(Y_{(i)}-y)(Y_{(i+1)}-y)<0\} + |r^{*}_{m+2}-r^{*}_{m+1}| + |r^{*}_{m+1}-r^{*}_{m}| - |r_{m+1}-r_{m}| \right],
\end{equation*}
where $r^{*}_{m+1}$ is the rank of $y$. 

Let 
\begin{align*}
\Delta_{n1} & = \frac{3}{n-1}\left\{ \sum_{i=1}^{n}\mathbf{1}[(Y_{i}-y)(Y_{N(i)}-y)<0] \right\} \\
& = \frac{3}{n-1}\left\{ n - \sum_{i=1}^{n}\mathbf{1}(Y_{i}<y)\mathbf{1}[Y_{N(i)}<y] - \sum_{i=1}^{n}\mathbf{1}(Y_{i}>y)\mathbf{1}[Y_{N(i)}>y]\right\},
\end{align*}
again by Lemma (9.10) of \cite{chatterjee}, we have
\begin{align*}
\lim_{n\rightarrow\infty}\mathbb{E}\{\mathbf{1}(Y_{i}<y)\mathbf{1}[Y_{N(i)}<y]\} & = \mathbb{E}_{X}[P^{2}(Y<y|X)],\\
\lim_{n\rightarrow\infty}\mathbb{E}\{\mathbf{1}(Y_{i}>y)\mathbf{1}[Y_{N(i)}>y]\} & = \mathbb{E}_{X}[P^{2}(Y>y|X)],
\end{align*}
therefore
\begin{align*}
\lim_{n\rightarrow\infty}\mathbb{E}(\Delta_{n1}) & = 6\mathbb{E}_{X}\left\{ P(Y>y|X)[1-P(Y>y|X)] \right\} \\
& = 6\mathbb{E}_{X}\left[ P(Y>y|X) \right] - 6\mathbb{E}_{X}\left[ P^{2}(Y>y|X) \right] \\
& = 6P(Y>y) - 6\mathbb{E}_{X}\left[ P^{2}(Y>y|X) \right]. 
\end{align*}

Next, we have
\begin{align*}
\frac{3}{n-1}|r_{m+1}-r_{m}| & = \frac{3n}{n-1}\frac{1}{n}|r_{m+1}-r_{m}| \\
& = \frac{3n}{n-1}|F_{Yn}(Y_{(m+1)}) - F_{Yn}(Y_{(m)})|. 
\end{align*}
Similar to $\Delta_{n1}$, we have
\begin{equation*}
\lim_{n\rightarrow\infty}\mathbb{E}\left[ \frac{3}{n-1}|r_{m+1}-r_{m}| \right ] =  \lim_{n\rightarrow\infty}\mathbb{E}\left[ 3|F_{Y}(Y_{(m+1)}) - F_{Y}(Y_{(m)})| \right ],
\end{equation*}
\begin{align*}
\lim_{n\rightarrow\infty}\mathbb{E}\left\{ \min[F_{Y}(Y_{(m+1)}), F_{Y}(Y_{(m)})] \right\} & = \lim_{n\rightarrow\infty}\mathbb{E}_{Y'}\left[ P(Y'< Y|X=X_{(m)}) P(Y'< Y|X=X_{(m+1)}) \right] \\
& = \mathbb{E}_{Y'}\left[ P^{2}(Y'< Y|X=x)\right],
\end{align*}
and
\begin{align*}
\lim_{n\rightarrow\infty}\mathbb{E}\left[ F_{Y}(Y_{(m)}) \right] & = \lim_{n\rightarrow\infty}\mathbb{E}_{Y'}\left[ P(Y'< Y|X=X_{(m)}) \right] \\
& = \mathbb{E}_{Y'}\left[ P(Y'< Y|X=x)\right].
\end{align*}

Furthermore, we have $\lim_{n\rightarrow\infty}\mathbb{E}\left[ F_{Y}(Y_{(m+1)}) \right] = \mathbb{E}_{Y'}\left[ P(Y'< Y|X=x) \right]$ and 
\begin{align*}
\lim_{n\rightarrow\infty}\mathbb{E}(\Delta_{n2})& = \lim_{n\rightarrow\infty}\mathbb{E}\left[ 3|F_{Y}(Y_{(m+1)}) - F_{Y}(Y_{(m)})| \right ] \\
& = \lim_{n\rightarrow\infty}3\mathbb{E}\left\{ F_{Y}(Y_{(m+1)}) + F_{Y}(Y_{(m)}) - 2\min[F_{Y}(Y_{(m+1)}), F_{Y}(Y_{(m)})] \right\} \\
& = 6\mathbb{E}_{Y'}\left[ P(Y'< Y|X=x) \right] - 6\mathbb{E}_{Y'}\left[ P^{2}(Y'<Y|X=x) \right]
\end{align*}

Similarly, for $\Delta_{n3}$, we have
\begin{align*}
\lim_{n\rightarrow\infty}\mathbb{E}(\Delta_{n3}) & =  6\mathbb{E}_{Y'}\left[ P(Y'< Y|X=x) \right] + 6\mathbb{E}_{Y'}\left[ P(Y'< y) \right] - 12\mathbb{E}_{Y'}\left[ P(Y'< Y, Y'<y|X=x) \right] \\
& = 6\mathbb{E}_{Y'}\left[ P(Y'< Y|X=x) \right] + 6P(Y< y) - 12\mathbb{E}_{Y'}\left[ P(Y'< Y, Y'<y|X=x) \right]
\end{align*}

Summarizing the results above, we have
\begin{align*}
\mbox{AESF}[(x,~ y),~ \xi,~ F] = & - 12\mathbb{E}_{Y'}\mathbb{E}_{X}\left[ P^{2}(Y'<Y|X) \right] + 6\mathbb{E}_{X}\left[ P^{2}(Y>y|X) \right]- 6\mathbb{E}_{Y'}\left[ P^{2}(Y'<Y|X=x) \right] \\
& + 12\mathbb{E}_{Y'}\left[ P(Y'< Y, Y'<y|X=x) \right],
\end{align*}
which is uniformly bounded.

As illustrative examples, we present in Figure 4 below the $AESF$ of Chatterjee's correlation under three correlative patterns: (A) linear (B) quadratic and (C) sinusoid. Let $Z\sim N(0 ,1)$ and $Z\perp X$, the three settings are constructed as follows
\begin{itemize}
\item[A.] $X\sim N(0, 1)$, $Y = 0.7X + \sqrt{1-0.7^2}Z$.
\item[B.]  $X\sim U(-10, 10)$, $Y = X^2 + \sqrt{10}Z$.
\item[C.]  $X\sim U(-1, 1)$, $Y = \cos(2\pi X) + 0.5Z$.
\end{itemize}
 
\begin{center}
[Figure 4 about here]
\end{center}

\section{Discussion and conclusions}
In this paper, we introduced a new robustness measure, $AESF$, which is defined as the limit of the expected sensitivity function. The $AESF$ bridges the gap between sensitivity and influence functions, exhibiting convenient properties for both. Notably, it shares equivalence with influence functions for many common estimators such as mean, variance, and Spearman's correlation and Kendall's correlation. The $AESF$ possesses several key advantages. First, unlike the sensitivity function which is generally random even when sample size approaches infinity, the $AESF$ is always deterministic, which is a desired property for robustness measures. Second, compared to the influence function based on G\^{a}teaux derivative, the $AESF$ are often more mathematically manageable, particularly for complex statistical functionals. We illustrate this with Chatterjee's correlation and Dette-Siburg-Stoimenov's dependence measure as an example. Furthermore, one may use this new metric to redefine some robustness concepts such as gross-error sensitivity and B-robustness, further enriching the understanding of statistical stability.

While Theorems 1 and Corollary 1 establish sufficient conditions for equivalence between $SF$, $AESF$, and $IF$ for a broad class of functionals and estimators, a deeper understanding requires a sufficient and necessary condition. The intricate relationship between $IF$ and $SF$ has been extensively studied. Boos \& Serfling (1980) demonstrated that under Fr\'{e}chet differentiability, $SF$ consistently estimates $IF$ asymptotically with normality. Subsequently, Cuevas \& Romo (1995) established the asymptotic validity of bootstrap confidence bands for $SF$ as an estimator of $IF$ using Gill's generalized method for Hadamard differentiable operators. However, these results hinge on pre-defined sufficient conditions. Achieving a sufficient and necessary condition presents a significant theoretical challenge, hence we leave it as an open question for future investigation.

The proposed robustness metric has far-reaching potential, in addition to the nonparametric correlation estimators showcased in this work. For instance, it readily applies to the robustness of diverse statistical and machine learning models, even without requiring the explicit form of their underlying functional. This versatility grants $AESF$ broad applicability within the realm of model robustness assessment.

\section*{Acknowledgement}
\noindent
The work was supported by an NSF DBI Biology Integration Institute (BII) grant (award no. 2119968; PI-Ceballos).

\section*{Appendix: Proof of Theorem 1}
\begin{proof}
We first derive $\mbox{IF}(x,~ R,~ F)$. Let $E_{g} = \int g(t) dF(t)$, we have
\begin{align*}
\mbox{IF}(x,~ R,~ F) & = \lim_{\epsilon\rightarrow 0}\frac{\phi[\epsilon g(x) + (1-\epsilon) E_{g}] - \phi[E_{g}]}{\epsilon} \\
& = [g(x) - E_{g}] \lim_{\epsilon\rightarrow 0}\frac{\phi\{E_{g} + \epsilon[g(x) - E_{g}]\} - \phi[E_{g}]}{\epsilon[g(x) - E_{g}]} \\
& = [g(x) - E_{g}] \phi^{'}[E_{g}].
\end{align*}
Next we show for any $x$, $\mbox{SF}(x, ~R_{n}, ~F_{n}) \xrightarrow[]{a.s.}[g(x) - E_{g}] \phi^{'}[E_{g}]$. Let $(X_{1},~ ...,~ X_{n})$ be i.i.d. samples, first we have
\begin{align*}
\mbox{SF}(x, ~R_{n}, ~F_{n}) & = \frac{\phi\left\{\frac{1}{n+1}\left[g(x) + \sum_{i=1}^{n}g(X_{i})\right]\right\} - \phi\left[ \frac{1}{n}\sum_{i=1}^{n}g(X_{i}) \right]}{\frac{1}{n+1}} \\
& = \left[ g(x) - \frac{1}{n}\sum_{i=1}^{n}g(X_{i}) \right] \frac{\phi\left\{\frac{1}{n}\sum_{i=1}^{n}g(X_{i}) + \frac{1}{n+1}\left[g(x) - \frac{1}{n}\sum_{i=1}^{n}g(X_{i}) \right]\right\} - \phi\left[ \frac{1}{n}\sum_{i=1}^{n}g(X_{i}) \right]}{\frac{1}{n+1}\left[ g(x) - \frac{1}{n}\sum_{i=1}^{n}g(X_{i}) \right]}.
\end{align*}
By the continuity and boundedness of $\phi^{'}$, for any $\delta>0$, there exists $\eta>0$, such that $|\phi^{'}(t)-\phi^{'}(E_{g})|<\delta$ for $E_{g}-\eta < t < E_{g}+\eta$. By SLLN, we have 
$$g(x) - \frac{1}{n}\sum_{i=1}^{n}g(X_{i})\xrightarrow[]{a.s.}g(x) - E_{g}$$ 
and 
$$\frac{1}{n+1}\left[ g(x) - \frac{1}{n}\sum_{i=1}^{n}g(X_{i}) \right]\xrightarrow[]{a.s.}0.$$
Therefore there exists $N_{1}$, such that for $n>N_{1}$, $|\sum_{i=1}^{n}g(X_{i})/n-E_{g}|<\eta/2$ almost surely. Also, there exists $N_{2}$, such that $n>N_{2}$, $|\frac{1}{n+1}\left[ g(x) - \frac{1}{n}\sum_{i=1}^{n}g(X_{i}) \right]|<\eta/2$ almost surely. Thus by the continuity and boundedness of $\phi^{'}$, for $n>\max\{N_{1}, N_{2}\}$, we have 
\begin{equation*}
\left|\frac{1}{n}\sum_{i=1}^{n}g(X_{i}) + \frac{1}{n+1}\left[g(x) - \frac{1}{n}\sum_{i=1}^{n}g(X_{i}) - E_{g}\right]\right|<\eta, ~ a.s.
\end{equation*}
and 
\begin{equation*}
\left|\frac{\phi\left\{\frac{1}{n}\sum_{i=1}^{n}g(X_{i}) + \frac{1}{n+1}\left[g(x) - \frac{1}{n}\sum_{i=1}^{n}g(X_{i}) \right]\right\} - \phi\left[ \frac{1}{n}\sum_{i=1}^{n}g(X_{i}) \right]}{\frac{1}{n+1}\left[ g(x) - \frac{1}{n}\sum_{i=1}^{n}g(X_{i}) \right]}-\phi^{'}(E_{g})\right|<\delta, ~a.s.
\end{equation*}
Therefore for any $x$, $\mbox{SF}(x, ~R_{n}, ~F_{n}) \xrightarrow[]{a.s.}[g(x) - E_{g}] \phi^{'}[E_{g}]$. 

Finally, we show that $\mbox{AESF}(x,~ R,~ F) = [g(x) - E_{g}] \phi^{'}[E_{g}]$. Recall $\mbox{AESF}(x,~ R,~ F) = \lim_{n\rightarrow\infty}\mathbb{E}[\mbox{SF}(x, ~R_{n}, ~F_{n})]$. By the boundedness of $\phi^{'}$, there exists $M>0$, such that $|\mbox{SF}(x, ~R_{n}, ~F_{n})|<M\left|g(x) - \frac{1}{n}\sum_{i=1}^{n}g(X_{i}) \right|$. As $\mathbb{E}\left|g(x) - \frac{1}{n}\sum_{i=1}^{n}g(X_{i}) \right| \leq |g(x)| + \int |g(t)| dF(t) < \infty$, by Lebesgue's Dominated Convergence Theorem, we have $\mbox{AESF}(x,~ R,~ F) = \mathbb{E}[\lim_{n\rightarrow\infty}\mbox{SF}(x, ~R_{n}, ~F_{n})]$. Since $\mbox{SF}(x, ~R_{n}, ~F_{n})]$ is uniformly bounded, the almost sure convergence also implies convergence in expectation, thus $\mathbb{E}[\lim_{n\rightarrow\infty}\mbox{SF}(x, ~R_{n}, ~F_{n})] = [g(x) - E_{g}] \phi^{'}[E_{g}].$ This completes the proof.
\end{proof}

\newpage
\section*{Figures}
\begin{figure}[!htbp]
\begin{center}
\includegraphics[scale=0.5]{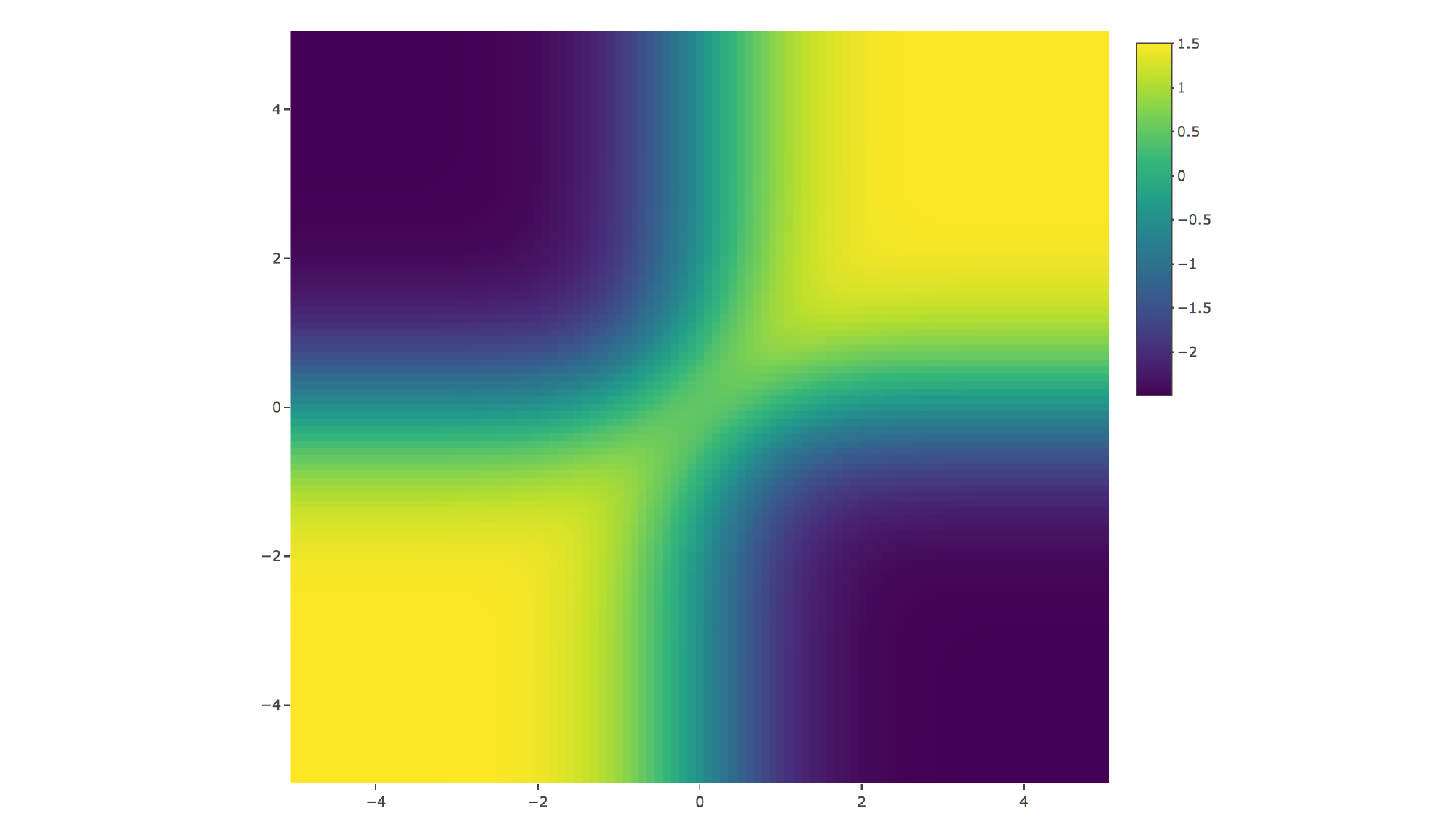}
\end{center}
\caption{The $AESF$ of Kendall's correlation for bivariate normal distribution with $\rho = 0.7$.
}
\end{figure}

\newpage
\begin{figure}[!htbp]
\begin{center}
\includegraphics[scale=0.5]{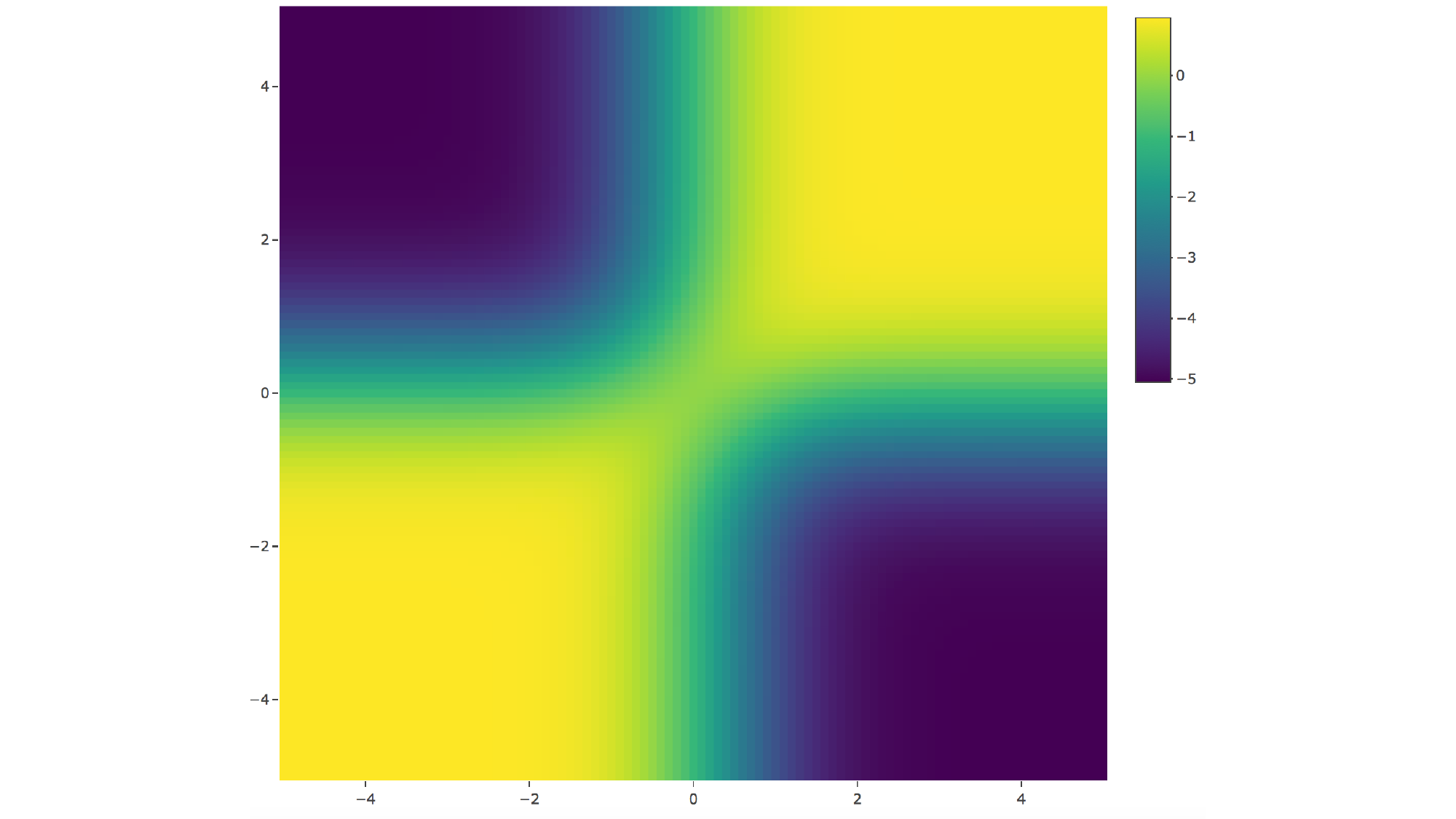}
\end{center}
\caption{The $AESF$ of Spearman's correlation for bivariate normal distribution with $\rho = 0.7$.
}
\end{figure}

\newpage
\begin{figure}[!htbp]
\begin{center}
\includegraphics[scale=0.5]{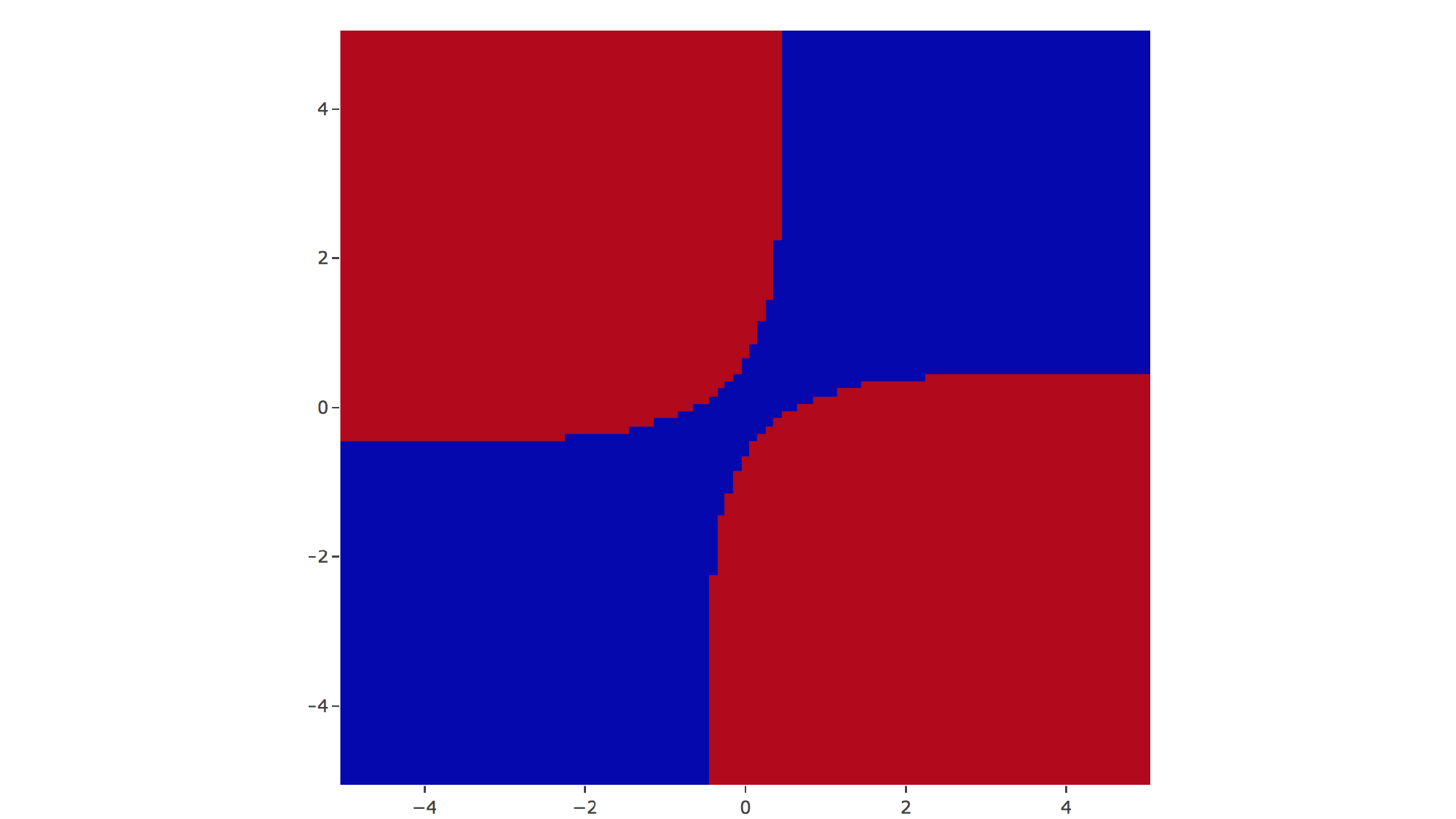}
\end{center}
\caption{Comparison of Kendall's and Spearman's correlations in $AESF$, where red highlights areas where Kendall's correlation is more robust (smaller absolute $AESF$), while blue indicate regions where Spearman's correlation is more robust.
}
\end{figure}

\newpage
\begin{figure}[!htbp]
\begin{center}
\includegraphics[scale=0.52]{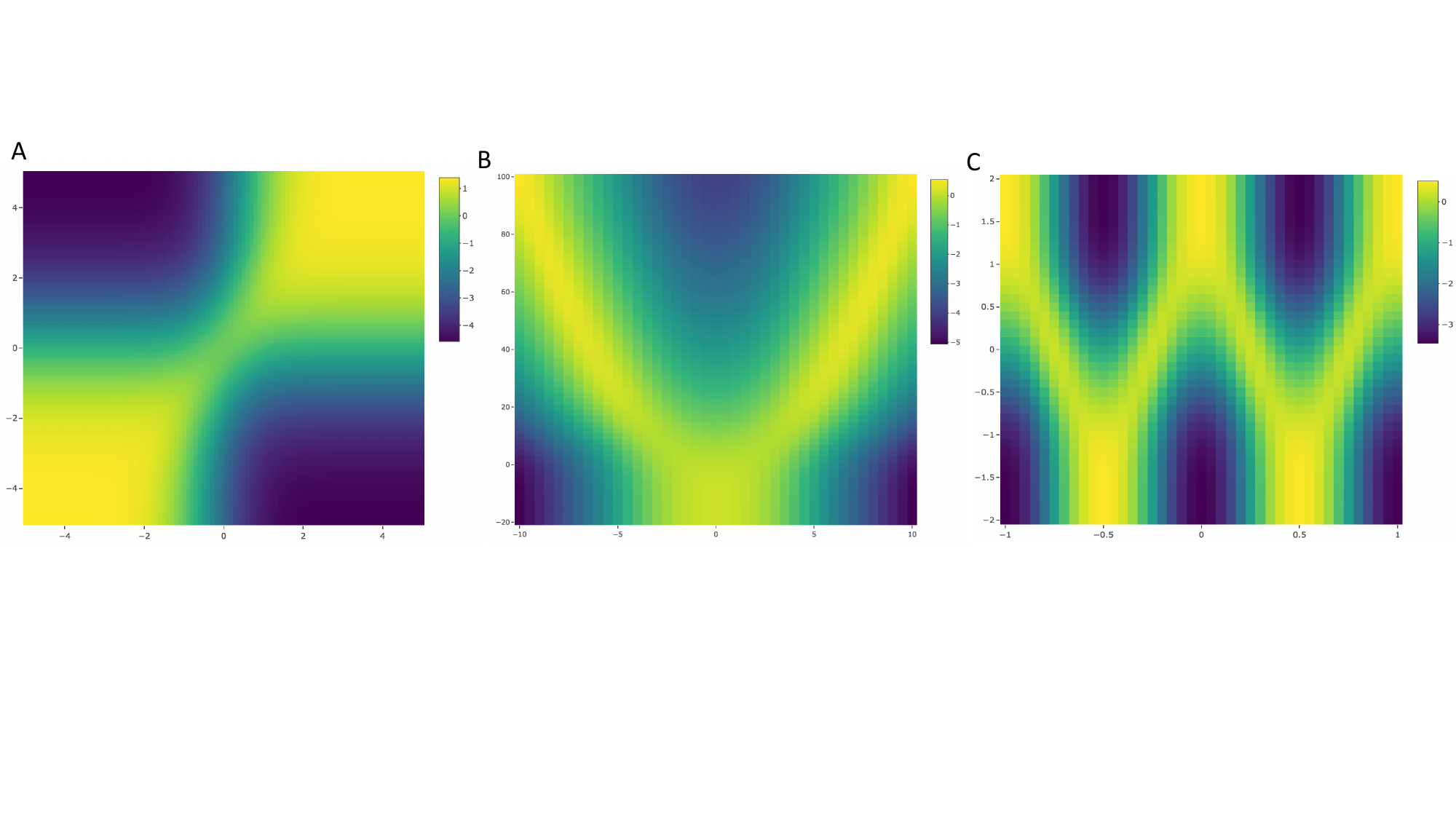}
\end{center}
\caption{The $AESF$ of Chatterjee's correlation for (A) linear (B) quadratic and (C) sinusoid, under Gaussian random noise.
}
\end{figure}

\end{document}